\documentclass[preprint2,longabstract]{aastex}

\newcommand{\ef}{\Psi}

\newcommand{\xmm}{XMM-{\em Newton}}

\newcounter{ion}

\def \etal   {\hbox{et~al.\/}}

\shorttitle{X-ray Light Curve Analysis of WR~6}
\shortauthors{Ignace et al.}

\begin{document}
\title{The {\em XMM-Newton} EPIC X-ray Light Curve Analysis of 
WR~6\footnote{Based on observations obtained with XMM-Newton, an ESA
science mission with instruments and contributions directly funded
by ESA Member States and NASA.}}

\author{R. Ignace}
\affil{Department of Physics and Astronomy, East Tennessee State 
University, Johnson City, TN 37663, USA\\
\email{ignace@etsu.edu}}
\author{K.~G. Gayley}
\affil{Department of Physics and Astronomy, University of Iowa, 
Iowa City, IA 52245, USA}
\author{W.-R. Hamann} 
\affil{Institute for Physics and Astronomy, University Potsdam, 
14476 Potsdam, Germany}
\author{D.~P. Huenemoerder}
\affil{Massachusetts Institute of Technology, Kavli Institute 
for Astrophysics and Space Research, 70 Vassar St., Cambridge, 
MA 02139, USA}
\author{L.~M. Oskinova}
\affil{Institute for Physics and Astronomy, University Potsdam, 
14476 Potsdam, Germany}
\author{A.~M.~T. Pollock}
\affil{European Space Agency XMM-Newton Science Operations Centre, 
European Space Astronomy Centre, Apartado 78, 
Villanueva de la Ca\~nada, 28691 Madrid, Spain}
\author{M. McFall}
\affil{Department of Physics, 191 W.\ Woodruff Ave, Ohio State University,
Columbus, OH 43210, USA
}

\begin{abstract}    

We obtained four pointings of over 100~ks each of the well-studied
Wolf-Rayet star WR 6 with the XMM-Newton satellite. With a first
paper emphasizing the results of spectral analysis, this follow-up
highlights the X-ray variability clearly detected in all four
pointings. However, phased light curves fail to confirm obvious
cyclic behavior on the well-established 3.766~d period widely found
at longer wavelengths. The data are of such quality that we were
able to conduct a search for “event clustering” in the arrival times
of X-ray photons. However, we fail to detect any such clustering.
One possibility is that X-rays are generated in a stationary shock
structure. In this context we favor a co-rotating interaction region
(CIR) and present a phenomenological model for X-rays from a CIR
structure. We show that a CIR has the potential to account
simultaneously for the X-ray variability and constraints provided
by the spectral analysis. Ultimately, the viability of the CIR model
will require both intermittent long-term X-ray monitoring of WR 6
and better physical models of CIR X-ray production at large radii
in stellar winds.  

\end{abstract}


\section{Introduction}

X-ray emission from massive stars continues to demonstrate its
importance for understanding these objects (e.g., G\"{u}del \& Naz\'{e}
2009).  In particular, X-ray generation can be associated with
nonthermal processes, like particle acceleration, or with irreversible
thermal processes, like shocked flows.  The hypersonic laboratory
afforded us by these winds could yield any of these emission types,
depending on the nature of the winds and the phenomena they support.
Thus the observed X-rays give us a unique window into the processes
that ultimately energize the interstellar medium and mediate galactic
evolution (e.g., Leitherer \etal\ 2010).  Previously known mechanisms
for generating X-rays from the winds of massive stars include
collisions in binary systems (e.g., Usov 1992; Stevens, Blondin,
\& Pollock 1992; Canto, Raga, \& Wilkin 1996; Walder \& Folini 2000;
Parkin \& Pittard 2008; Gayley 2009), shock production in magnetically
confined wind streams (Babel \& Montmerle 1997; Townsend, Owocki,
\& ud-Doula 2007; Li \etal\ 2008; Oskinova \etal\ 2011; Petit \etal\
2013; Ignace, Oskinova, \& Massa 2013), and time-dependent shocks
arising from inherent instabilities in line-driven winds (Lucy \&
White 1980; Lucy 1982; Owocki, Castor, \& Rybicki 1988; Feldmeier,
Puls, \& Pauldrach 1997; Dessart \& Owocki 2003).  The latter two
mechanisms arise in single stars, and came as a surprise when they
were originally detected (Seward \etal\ 1979; Harnden \etal\ 1979).
The shocks from the line-driven instability (LDI) are expected to
be somewhat weaker, perhaps in the characteristic temperature range
$kT \sim 0.1-1$~keV, than the strong shocks from magnetically
confined winds.  However, it should be noted that both of these
mechanisms operate relatively close to the star, as line driving
occurs where the wind is accelerating and magnetic channeling
requires strong fields.  In this paper, we will comment on X-ray
generation that is inferred to originate well beyond the
acceleration zone of the wind.

We note that in the roughly four decades since the discovery of
X-rays from massive stars, the quality of the information has
increased markedly.  Modern X-ray observatories like the {\em
XMM-Newton} and {\em Chandra} telescopes have larger collecting
area and better spectral resolution than ever before (e.g., Jansen
\etal\ 2001; Weisskopf \etal\ 2002).  Stellar winds are an example
of an area that have been significantly impacted by these observational
advances in the X-ray band.  The target discussed here is a member
of the class of Wolf-Rayet (WR) stars, a relatively rare type of
massive star.  Although rare, the WR~stars command significant
attention by virtue of their extreme winds and evolved states.

%And our knowledge about the diversity of massive stars has changed
%as well.  It is known now that some massive stars can harbor
%significant surface magnetic fields (e.g., Hubrig \etal\ 2008;
%Donati \& Landstreet 2009; Hubrig \etal\ 2011; Hubrig \etal\ 2013;
%Grunhut \& Wade 2013), extensive magnetospheres that substantially
%modify their circumstellar environment and produce hot, X-ray
%emitting plasma.

A topic of particular interest has been the production of X-ray
emissions in the winds of single massive stars, like our source
WR~6 (also EZ~CMa and HD~50896).  Here we present the second paper
reporting on 439~ks of {\em XMM-Newton} time.  Whereas the first
report by Oskinova \etal\ (2012; hereafter Paper~I) emphasized
information content provided by {\em XMM-Newton} spectroscopy of
WR~6, here the focus is on understanding the star's wind structure
through an analysis of the X-ray variability that it displays.

Wolf-Rayet (WR) stars are generally accepted to be a phase of massive
star evolution prior to termination as core-collapse supernova
(Lamers \etal\ 1991; Langer 2012).  Hydrogen is observed at much
lower abundances than solar, or may even be altogether absent.  The
WR~stars come in three principal subgroups: nitrogen-rich, carbon-rich,
and oxygen-rich, all of which are helium-rich.  Our target, WR~6,
is a WN4 star, indicating that it is an early-type star of the
nitrogen-rich category (Hamann, Koesterke, \& Wessolowski 1995;
Hamann, Gr\"{a}fener, \& Liermann 2006).  The winds of WR stars are
also expected to suffer from the LDI mechanism (Gayley \& Owocki
1995) and should thus emit X-rays similar to their O-type progenitors.
However, the WR winds tend to be more massive than for O~stars.  In
general the two types of wind have similar wind terminal speeds,
$v_\infty$, in the 1,000--3,000 km s$^{-1}$ range, but WR stars
have wind mass-loss rates $\dot{M}$ that are greater by up to an
order of magnitude relative to O stars of similar luminosity.  As
a result the winds are far more dense than O~star winds (e.g.,
Abbott \etal\ 1986; Bieging, Abbott, \& Churchwell 1989).  The large
wind densities and metallicity of WR winds make them quite opaque
for the X-rays. The large wind opacity was invoked in Oskinova
\etal\ (2003) to explain the apparent lack of X-ray emitting single
WC-type stars.

Perhaps the key result of the first paper is the measurement of
resolved X-ray line profile shapes and properties that are consistent
with X-ray emission that emerges from the wind at large radius (of
order $10^2 - 10^3 R_\ast$ in the wind, depending on the wavelength
of observation).  One evidence for this is the
nominal $f/i$ ratios observed in He-like triplet species (Gabriel
\& Jordan 1969; Blumenthal, Drake, \& Tucker 1972).  Here ``f''
refers to the forbidden component and ``i'' the intercombination
one.  The ratio of fluxes in these emission lines is a diagnostic
of pumping from the upper level of the ``i'' line to the ``f'' line.
In the absence of such pumping, the ratio has a value predicted by
intrinsic branching ratios (e.g., Porquet \etal\ 2001), but the
ratio can be reduced by either collisions or radiative excitation.
The former is only important at high densities that are either not
present in winds, or present only at depths from which X-rays may
have a difficult time emerging, so is not considered here.  Radiative
pumping is of more potentially ubiquitous importance in the UV-bright
circumstellar environment of massive stars, and then the $f/i$ ratio
becomes a diagnostic of the dilution of the stellar continuum,
namely {\em where} in the wind X-ray emissions are formed (Waldron
\& Cassinelli 2001; Cassinelli
\etal\ 2001).

For WR~6 the $f/i$ ratios are close to their intrinsic and unpumped
values, placing lower limits in the wind for the radius of the X-ray
emission (see Paper~I).  In addition, the line profiles are asymmetric
in conformance with expectations for X-ray production that is
distributed throughout, and substantially photoabsorbed by, a dense
and metal-rich WR wind (Ignace 2001).  This stands in considerable
contrast to O~star winds, where line profile shapes are often more
symmetric in appearance, and $f/i$ ratios tend to show anomalous
values that are consistent with UV pumping and therefore proximity
of the hot X-ray emitting plasma to the stellar photospheres (e.g.,
Waldron \& Cassinelli 2007).

Although the high degree of photoabsorption from the dense WR~wind implies
that observable emission would need to come from a large radius,
possibly even $10^3$ stellar radii, the challenge for these results
is to explain how any X-ray generating process in a single star could
operate efficiently so far from the region where line driving and
potentially strong magnetic fields could be present.  In addition,
WR~6 has a well-known ``clock'' in that it shows polarimetric and
spectroscopic variability on a period of 3.766 days (e.g., Firmani
\etal\ 1980), yet no binary companion has been detected.  Moreover,
the X-rays seem too soft and of too low a luminosity to be associated
with a compact companion (e.g., Morel, St-Louis, \& Marchenko 1997).
Based on the enhanced hardness of the emission in WR~6 compared
to O~stars, Skinner \etal\
(2002) have suggested that a low-mass non-degenerate companion might
explain the X-rays.  However, for a mass of $M_\ast \sim 30M_\odot$
and a radius $R_\ast \sim 1 R_\odot$, a 3.766~d period corresponds to
a semi-major axis of only $30R_\ast$, which is about an order of
magnitude smaller than the radius where optical depth unity is achieved
by wind photoabsorption at 1~keV.  Perhaps some hard emission from a
wind collision onto a companion could emerge owing to the fact that the
photoabsorption opacity declines steeply with increasing photon
energy, but certainly little or no
soft emission would escape.  Also, the spectroscopic variability does
not show the strict phase coherence from cycle to cycle that might be
expected from a binary companion, but is consistent with a rotating star
with stochastically varying features.

Given that the evidence so far favors a single-star hypothesis for
WR~6, this presents difficulties in accounting for the observed
X-rays.  Theorists have appealed to large-scale clumping and porosity
in stellar winds as a geometrical effect to allow for easier escape
of X-ray photons (Feldmeier, Oskinova, \& Hamann 2003; Owocki \&
Cohen 2006; Oskinova, Hamann, \& Feldmeier 2007; Sundqvist \etal\
2012).  The evidence for clumping in massive star winds is certainly
voluminous (e.g., Hillier 1991; Moffat \& Robert 1994; Hamann \&
Koesterke 1998), and clumping ameliorates the X-ray emission-line
profile asymmetries that result from smooth, non-clumped wind
considerations.  

Clumping on the smaller scale of the photon mean-free path, sometimes
termed ``microclumping,'' provides no real help in terms of X-ray
photon escape, because it serves only to enhance emissivity at fixed
mass-loss rate.  Given that optically thin photoabsorption is linear
in density, it scales the same as does the mass-loss rate itself,
and is generally already included in mass-loss estimates.

This paper reports on the analysis and interpretation of X-ray
variability detected from WR~6.  Section~\ref{sec:obs} provides a
brief review of the dataset, which has been discussed more thoroughly
in Paper~I.  Section~\ref{sec:anal} presents an analysis of the
variability data, incorporating not only the recent {\em XMM-Newton}
pointings, but also all prior pointings from archival data.
Section~\ref{sec:model} presents a discussion of the results in
terms of potential causes, and explores the possibility of accounting
for the observed variability in terms of a co-rotating interaction
region (CIR) model.  Concluding remarks are given in
Section~\ref{sec:conc}.

\section{Observations}
\label{sec:obs}

The X-ray data on WR\,6  were taken with the X-Ray  Multi-Mirror
Satellite \xmm.  Its telescopes illuminate three different  instruments
which always operate simultaneously: RGS is a Reflection Grating
Spectrometer, achieving a spectral resolution of 0.07\,\AA; RGS is
not sensitive for wavelengths shorter than 5\,\AA. The other focal
instruments MOS and PN cover the shorter wavelengths; their spectral
resolution is modest ($E/\Delta E \approx 20 - 50$).  

\begin{figure}[t]
%\hspace{-1.5em}\includegraphics[width=.76\columnwidth, angle=-90]{boo1.ps}
\hspace{-1.5em}\includegraphics[width=.76\columnwidth, angle=-90]{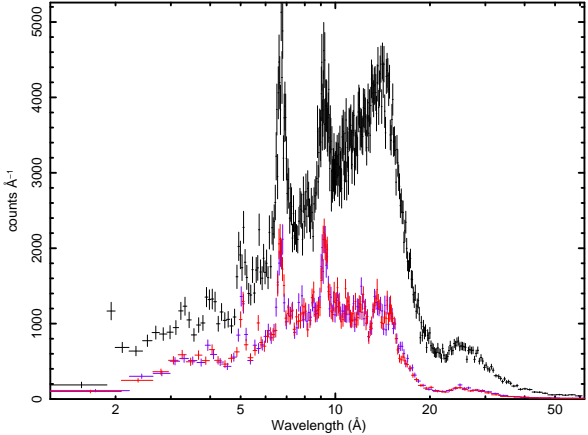}
\caption{ {\em XMM-Newton} PN (upper points in black), and MOS1 and
MOS2 (lower points in red and blue) spectra of WR\,6 obtained on 6
Nov 2010. The error bars correspond to 10$\sigma$.  Note that the
spectra are displayed as counts per Angstrom in a linear scale
against wavelength in a logarithmic scale.  Please refer to Paper~I
for a thorough discussion of spectral features.}
\label{fig1} 
\end{figure}

The data were obtained at four epochs in 2010 (Oct.\,11 and 13,
Nov.\,4 and 6; see Tab.~\ref{tab1}). The total exposure time of 439~ks
was split into four individual parts.  Each exposure
was approximately 30~hr in duration.  The observations were not strongly
affected by soft proton background flares.  Our data reduction
involved standard procedures of the \xmm\ Science Analysis System
v.10.0.  Figure~\ref{fig1} displays the observed EPIC spectrum for
the full exposure provided during the first of our four pointings.
The three colors are for instruments PN, MOS1, and MOS2.  Spectral
analysis was the focus of Paper~I.  Here we explore the implications
of observed variable X-ray emissions from WR~6.

In order to display, analyze, and discuss the X-ray variability of
WR~6, we have evaluated spectrum-integrated count rates from the
three EPIC instruments.  Given the stability of the EPIC-PN instrument,
a combined count rate $\dot{C}_{\rm T}$ was determined from a
simultaneous fit to the count rates in the three instruments assuming

\begin{equation}
\dot {\cal C}_{\rm T} = \dot{\cal C}_{\rm PN} = \frac{M(\dot{\cal C}_{\rm PN})}
	{M(\dot{\cal C}_{\rm M1})}\,\dot{\cal C}_{\rm M1} =
	\frac{M(\dot{\cal C}_{\rm PN})}{M(\dot{\cal C}_{\rm M2})}\,
	\dot{\cal C}_{\rm M2}.
\end{equation}

\noindent where $M(x)$ represents the median of a set of values in
an observation.
Minimization techniques were used on image frames for the three
independent detectors to obtain the best estimate of the combined count rate.
Background counts were modeled as well and treated independently
among the instruments.  The final count rate $\dot{C}_{\rm T}$
represents that value for the source that statistically produces
the most consistent results for all three instruments.  In this way
signal-to-noise is maximized to produce the most sensitive possible study
of variability in WR~6 from our dataset.

\begin{table}
\begin{center}
\caption{\xmm\ Observations of WR\,6 \label{tab1}}
\vspace{1em}
\renewcommand{\arraystretch}{1.2}
\begin{tabular}[h]{lcc}  \hline
\hline
Dataset ID & Date &  Duration  \\
            &        &  (ks)    \\
\hline
0652250501 & 2010-10-11 & 111 \\
0652250601 & 2010-10-13 & 105 \\
0652250701 & 2010-11-04 & 111 \\
0652250101 & 2010-11-06 & 112 \\
\hline
\end{tabular}
\end{center}
\end{table}

\section{Analysis}
\label{sec:anal}

Our analysis of X-ray variability from WR~6 takes three basic forms.
First, we consider event rate clustering, which seeks to determine
whether the arrival of X-ray photons are clustered in time.  Next we
make a formal evaluation of the variability in WR~6 by considering
the distribution of count rates from the star.  Finally, we present
light curves from the four pointings and phase these on the 3.766~d
period of WR~6.  We also consider how the source varies in different
energy bands of the spectrum.

\subsection{Event Rate Clustering}

Such a long exposure has allowed us to search for clustering of
X-ray detections in the event log of the detections.  Clustering
in the event log can provide a probe of spatially coherent shock
structures within the wind that produce the detected X-ray flux.
For example, consider a shock that develops instantaneously in the
wind flow and cools: the time-dependent X-ray emission would take
the form of a jump in counts followed by a period of decay
characteristic of the cooling time.  If the structure is substantially
extended in a lateral sense, there could even be a modification to
the temporal form of the signal (in both shape duration) owing to
finite light travel time effects.  This kind of ``pulse'' would
lead to a clustering of detected events in the event log.

Natually, we would expect in a clumped wind flow that there are many
of these shock events occuring throughout the flow.  If X-ray photons
were produced in an entirely random fashion, then we should expect a
Poisson distribution for the detected events from a stochastic process.
However, the physics of the shocks suggest that although the occurrence
of shock events may be random, the signals that they produce are not.

Each of the four pontings to WR~6 provided an exposure of $\sim 10^5$
seconds and a listing of source detection events yielding roughly
$\sim 60,000$ source counts per pointing for the PN detector.  To
test for clustering of events in photon arrival times, we conducted
the following experiment.  We chose a time interval $\Delta t$.
With a given value of $\Delta t$, we step through an event list for
one of the pointings.  For an event occurring at time $t_{\rm i}$
for the $i^{\rm th}$ event, we count the number of additional neighboring
events that fall in the interval $t_{\rm i} \pm \Delta t$.  This
is done for every event, and for a range of $\Delta t$ values.

\begin{figure}[t!]
%\hspace{-0.5em}\includegraphics[width=.75\columnwidth, angle=-90]{WR_fig.ps}
\hspace{-0.5em}\includegraphics[width=.75\columnwidth, angle=-90]{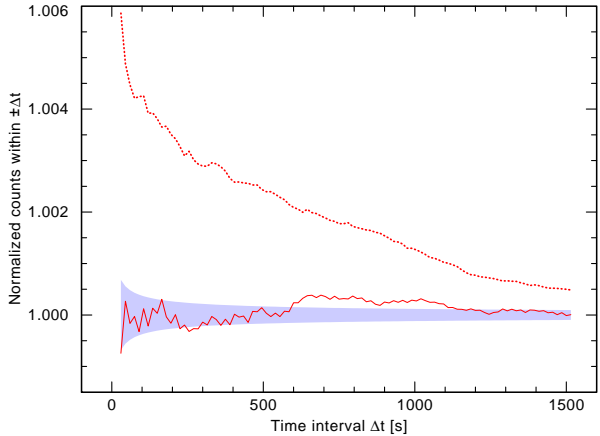}
\caption{
An event list with the $\sim 60,000$ source photons at the PN
detector has been extracted from the $\sim 100,000$\,s exposure for
the fourth pointing and analyzed for clustering as described in the
text. The data shown as the solid red line comply with completely
random, un-clustered events, which would give unity in the normalized
representation of the figure with one sigma of Poisson noise as
indicated (purple shaded area). For comparison, simulated data
assuming that $60,000$ photons were emitted in $10^4$ ``flares'',
randomly distributed over the exposure time, and each one decaying
exponentially with a time scale of 1,000\,s is displayed as the red
dotted line.  Despite the small number of only 6 photons per flare,
our sensitive test would reveal a very significant degree of
clustering.  }
\label{fig2} 
\end{figure}

Figure~\ref{fig2} displays the results of our experiment.  The solid
red line is based on the calculations for the data.  It represents
the number of events falling within $\pm \Delta t$ of a given event
at time $t_{\rm i}$.
These are normalized in such a way that the expectation from a Poisson
distribution would be unity.  The curve is plotted against the temporal
``window,'' $\Delta t$, in seconds.  The purplish band is the $1\sigma$
error band indicating the dispersion about the expected value of unity
for Poisson statistics.  As can be seen, the data are quite consistent
with pure Poisson noise:  most of the curve lies within or very near
the $1\sigma$ band.

As a test, we also experimented with simulated data.  We imagine that
X-rays are produced physically through shock events that we generically
refer to as ``flares''.  The occurrence of a shock generates X-ray
photons and events at the detector.  We assume that a total number of $N$
equally bright flares are randomly distributed over the exposure in time.
These flares are further assumed to cool exponentially, for which we
adopt a characteristic cooling time of $10^3$~s.  This rough cooling
time is based on an estimate of the density and clumping in the
wind.  The dotted red curve in Figure~\ref{fig2} represents application
of our event clustering diagnostic for a wind with $N = 10^4$ flares.
This curve lies well outside the $1\sigma$ band indicating that
clustering would have been easily detected under these conditions
if existing in the wind of WR~6.  The implication seems to be that
either there exist a very large number of ``flare'' events in the
wind of WR~6 to suppress detection of event clustering, or the
X-rays are produced in a stationary shock.  The latter is something
that we will explore further in Section~\ref{sec:model}.

\subsection{Examination of Source Count Rates}

\begin{figure}[t]
\centering
\includegraphics[width=1.0\columnwidth]{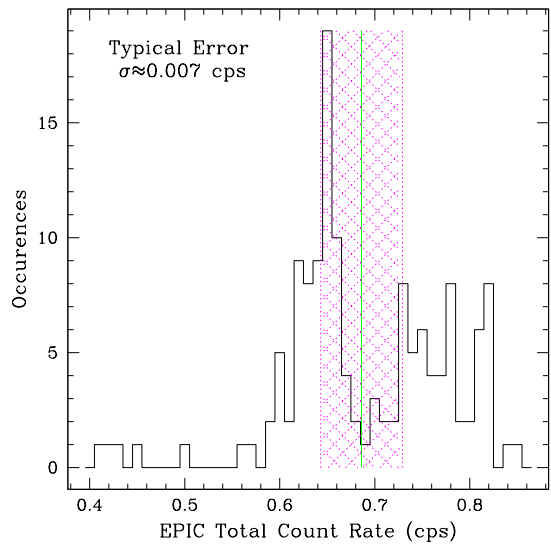}
\caption{
A histogram of the spectrum-integrated
count rates (i.e., all four bands together)
from the four new pointings using
the EPIC detector.  The vertical green line indicates the
error-weighted average count rate.  The magenta hatched region
is the $3\sigma_{\rm av}$ band about this average, for which
$\sigma_{\rm av}$ is the error in the mean count rate.
The binning interval is 0.1 cps with 140 count rate samples.
} 
\label{fig3} 
\end{figure}

Figure~\ref{fig3} displays a histogram of the total count rates obtained
from the PN, MOS1, and MOS2 instruments.  The count rates, in counts
per second (cps), from all four pointings have been binned at 0.1 cps
intervals to produce this figure of the frequency at which count rates
appear in the data for WR~6.  Note that the typical error in the count
rate is about 0.007 cps.

The vertical green line in the figure signifies the mean count rate.
Adopting notation that $\dot{{\cal C}}_{\rm i}$ is the total count rate
for the $i$th sample, the mean value is given by an error-weighted sum:

\begin{equation}
\langle \dot{{\cal C}} \rangle = \frac{\sum_i^N \dot{{\cal C}}_{\rm i}/
	\sigma^2_{\rm i}}{\sum_i^N 1/\sigma^2_{\rm i}},
\end{equation}

\noindent where $\sigma_{\rm i}$ is the error in the $i$th count rate.
The error in the average, $\sigma_{\rm av}$, is computed from

\begin{equation}
\sigma^2_{\rm av} = \frac{1}{N}\,\sum_i^N \sigma^2_{\rm i},
\end{equation}

\noindent where $N$ is the number of count rate samples.  

\begin{figure}[t]
\centering
\includegraphics[width=1.0\columnwidth]{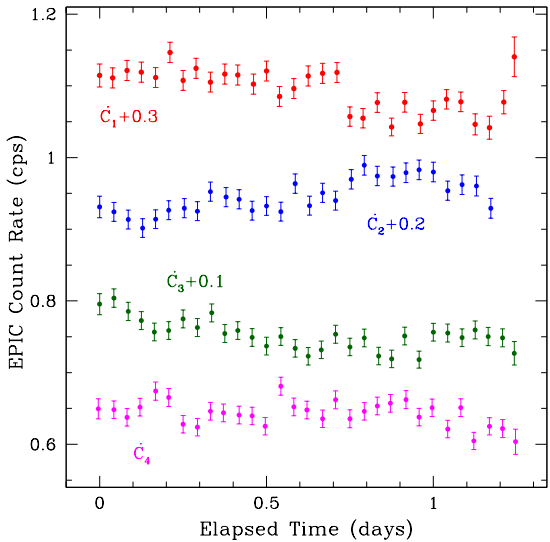}
\caption{
These are light curves for the four {\em XMM-Newton} pointings obtained
for WR~6 plotted in counts per second with elapsed time.  The total count rate
is from integrating the entire EPIC spectra, including PN, MOS1, and
MOS2, and averaging results for 3,600 second time bins.  Elapsed time
refers to days since the beginning each respective pointing.
The separate pointings have been shifted vertically in count rate
$\dot{\cal C}_{\rm T}$ as indicated.  The four are displayed with the first
at top, and the last at bottom.  The count rates include $1\sigma$
errors. }
\label{fig4} 
\end{figure}

The magenta-hatched region in Figure~\ref{fig4} is a band of
half-width $3\sigma_{\rm av}$.  For a non-varying source, one would
generally expect that over 99\% of the count rate samples would
fall within this band, for a normal distribution.  As can be seen,
the distribution of count rates is much broader than this hatched
region; indeed, it appears to be bimodal, with a grouping of lower
count rates at around 0.65~cps, and a less well-defined grouping
at higher values of around 0.77~cps.  It is clear that WR~6 displays
substantial X-ray variability at the 10--20 \% level over the
duration of our dataset.

\subsection{X-ray Light Curves}

Figure~\ref{fig4} shows light curves for our four separate pointings.
These are total count rates $\dot{\cal C}_{\rm k}$ for $k=1$ to 4, signifying
the sequence of pointings.  The four light curves have been vertically
shifted, as indicated, for clarity of viewing.  The horizontal is
elapsed time, in days, for each pointing separately.  In other words each
pointing is taken to begin at zero time.  The binning of counts
is over about 3,600~s for this figure.
Variability is evident.  Notable are trends that appear to persist
throughout some of the pointings.  For example, the second pointing
suggests a steadily increasing count rate; by contrast, the third one
indicates a declining one throughout the exposure.

\begin{figure}[t]
\centering
\includegraphics[width=1.0\columnwidth]{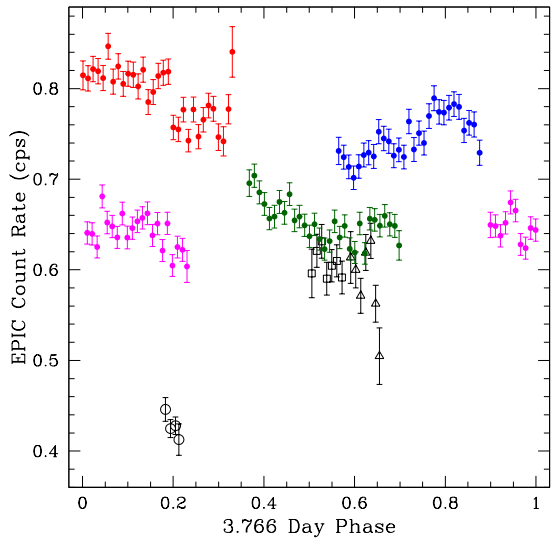}
\caption{
The count rates of Figure~\ref{fig4} are here displayed as phased
to the 3.766~d period of WR~6.  An arbitrary zero point
for the phasing was set to the beginning of the first of
the four new pointings obtained by the authors.  As in
Figure~\ref{fig4}, the first
pointing is shown in red; the second in blue; the third in
magenta; and the fourth in green.  Black points are for archival
data.  Although variability is evident, there is no coherent
X-ray light curve in terms of the 3.766~d period.
}  
\label{fig5} 
\end{figure}

Of particular interest for WR~6 is the well-known 3.766~d period
that appears to govern variability in this star's photometric,
polarimetric, line emissions (Duijsens \etal\ 1996; St-Louis \etal\
2009).  The period is discerned in lines from the UV, optical, and
IR.  We have taken the total EPIC count rate data and phased the
light curves on the 3.766~d period.  The phased light curve is shown
in Figure~\ref{fig5}.  Only data for our four pointings are shown
in color:  red for the first pointing, blue for the second, magenta
for the third, and finally green for the fourth; these are the same
colors as used in the preceding figure.  Additionally, the black
points refer to archival data.  The phasing of data arbitrarily
adopts the beginning of the pointing for the red dataset as zero
phase.  Note that the three measures shown as black circles near a
phase of 0.2 are archival data in which ``Thick'' EPIC filters were
used; Thick filter count rates tend to be about 20\% lower than for
the Medium filter.

Although the 3.766~d period is known to be stable over long timespans,
meaning that this is a timescale that has been consistently observed
over decades, there is no well-defined ephemeris for the variability.
Although the timescale persists, the ephemeris appears to drift between
epochs.  Conseqently, it comes as no surprise that the archival data fail
to line up with the more recent dataset.  But perhaps it is surprising
that the four pointings from the recent dataset also do not conform to
a coherent phased light curve.

The {\em XMM-Newton} is not capable of obtaining a single continuous 
439~ks light curve for a target source, and we did not request constrained
observations.  Even so, 439~ks is approximately four days of observing
time, roughly equal to the known period.  The four independent pointings
have substantial overlap within that cycle.  The second pointing (blue)
was obtained about 1 day after the first pointing (red) was completed.
The fourth pointing (green) was also obtained about 1 day following
the completion of the third one (magenta).  However, a time interval of
about 3 weeks separate these two pairs, corresponding to about 6 cycles
of the 3.766~d period.  Although the two separate pairs show similar
levels of variablity, the pairs are clearly shifted in terms of their
relative average count rates.  The error in a given sample count rate is
about 0.01 cps, whereas the separate pairs of pointings have average count
rates that are separated by roughly 0.1 cps.  This separation is the cause
for the bimodal appearance of total count rates seen in Figure~\ref{fig3}.

\begin{figure}[t!]
\centering
\includegraphics[width=1.0\columnwidth]{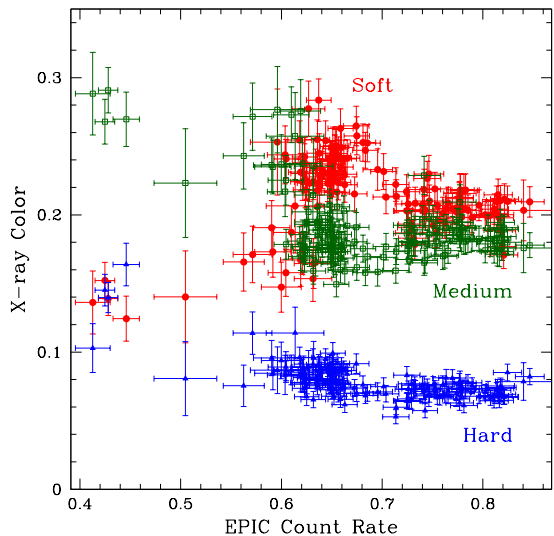}
\caption{
Similar to a color-magnitude plot, this figure shows three X-ray
colors plotted against total EPIC count rate.  The ``colors'' are
ratios of the count rates from different bands 1, 3, and 4 to that
of Band \#2, which has the highest count rate among the respective
bands.  The red points are ratios of count rates $\dot{\cal
C}_1/\dot{\cal C}_2$; green is for $\dot{\cal C}_3/\dot{\cal C}_2$,
and blue is for $\dot{\cal C}_4/\dot{\cal C}_2$.  The trends displayed
in this figure are discussed in the text.
} 
\label{fig6} 
\end{figure}

To further elucidate the nature of this variability, we have
formulated four energy bands to compare and contrast variability
in the soft and hard portions of the X-ray spectrum.  Table~\ref{tab2}
introduces the definitions of these bands.  The band count rates
are shown in Figure~\ref{fig6} in an analogue to a color-magnitude
diagram.  The abscissa is the total EPIC count rate.  The ordinate
is a ratio of band count rates such as $\dot{{\cal C}}_{\rm j}/
\dot{{\cal C}}_2$, for $j\ne 2$.  Count rates from Band \#2 were
chosen as the normalization because its values were largest, and
the errors smallest, plus it represents the majority of the counts
from the observed spectrum.  The red points are for $\dot{{\cal
C}}_1/\dot{{\cal C}}_2$ for the softest emission relative to Band
\#2; green is for $\dot{{\cal C}}_3/\dot{{\cal C}}_2$ labeled as
``Medium'', and blue is for $\dot{{\cal C}}_4/\dot{{\cal C}}_2$ for
the hardest emission.  Again, bimodality is evident in the appearance
of a pair of vertically shifted
groupings in total count rate within each relative
color (see Fig.~\ref{fig3}).
Ignoring the especially low count-rate data, there appear to be
some differences in the X-ray colors at the high ($\sim 0.8$ cps;
hereafter ``bright'') and low ($\sim 0.65$ cps; hereafter ``dim'')
count-rate pointings.  These are summarized as follows:

\begin{itemize}

\item Relative to Band \#2, all other Bands (1, 3, and 4) are less
bright, hence $\dot{\cal C}_{\rm j}/\dot{\cal C}_2 < 1$.

\item The Soft color (red points) is the strongest of the three
shown, and the Hard color (blue points) is the weakest.

\item The Medium color level is similar for both the bright
and dim states.  However, it appears that the Soft color level
is noticably increased for the dim state as compared to the bright
one.  The same may be true for the Hard color, but if so, the change
is less pronounced.

\item There are a handful of ``spurious'' points at low count rates,
below about 0.6 cps.  We have ignored these in the discussion of
trends above.  However, it does seem that there have been instances
when the X-ray brightness has at times dropped by nearly a factor
of 2 relative to the observed bright state.  At those times the
Medium color appears to be greatest, and the Soft and Hard color
levels are commensurate.  In particular, it seems that the Soft
color has dropped by a bit less than half, whereas the hard one has
increased by a bit less than two.  Such a spectral distribution 
could indicate a change in how the X-rays are generated, or might
indicate a change in the amount of photoabsorptive absorption, perhaps
the result of increased mass loss from the star.

\end{itemize}

\begin{table}
\begin{center}
\caption{Energy Bands	\label{tab2}}
\begin{tabular}{cc}
\hline\hline Band & Energy Interval \\ 
 & (keV) \\ \hline
\#1 & 0.3--0.6 \\
\#2 & 0.6--1.7 \\
\#3 & 1.7--2.7 \\
\#4 & 2.7--7.0 \\ \hline
\end{tabular}
\end{center}
\end{table}

There is no doubt that WR~6 displays fairly significant variability
of its X-ray emissions.  In general,
variability is a common property of WR stars. Among rigorously
monitored  WN stars, 40\%\ show optical variability similar to WR\,6
Chene \etal\ (2011).  Our high-resolution X-ray spectra from Paper~I
indicate that the X-ray emitting plasma moves at about the same velocity
as the cool wind, and the X-ray line blue-shifts do not change with time.
Also, the X-ray emission-line spectrum is compatible with the WN star
abundances.  All of these facts are consistent with an interpretation
of the X-rays as being endemic to a single-star wind.  Nevertheless,
whenever hard X-rays are generated at a large distance from a star, one
must carefully eliminate the possibility of binarity before looking
to more exotic explanations.  But as mentioned previously, an orbital
period of 3.766 days corresponds to an orbital semi-major axis for
a low-mass companion of only $30 R_\ast$, much less than the radius
of optical depth unity in photoabsorption predicted from models (cf.\
Paper~I) of about $10^2 R_\ast$ at an energy of 1~keV.  By contrast if
a low-mass companion were situated out at a distance of $10^2R_\ast$,
the orbital period $P_{\rm orb}$ would be at least 20~days,
considerably longer than the 3.766~d ``clock'' inherent to WR~6.

Still, the challenge to explain X-ray emissions at large radii remains.
Ignoring the clock in WR~6, if binarity is to be a plausible explanation
for the observed X-rays, it seems that the companion would have to
be in a somewhat large orbit of at least 100's of $R_\ast$, with a
period of a year.  Our viewing perspective would likely need to be
more pole-on than edge-on, to prevent drastic orbital modulation of the
X-ray luminosity.  It could be somewhat eccentric to produce longer term
variations of the X-ray luminosity that are not too great in amplitude.
Short-term variations (at the level of a day) in X-rays could then arise
as an effect of instabilities in the structure of the colliding wind
bow shock (e.g., Pittard \& Stevens 1997), or as a result of a small
number of very large wind clumps encountering the bow-shock region (e.g.,
Walder \& Folini 2002).

A serious difficulty in ruling out either of those scenarios is
that the wind flow time across a scale of $10^2 - 10^3 R_\ast$ is
half a day to days in duration, about the same as the UV periodicity.
But the UV periodicity comes from much deeper in the wind where the
flow time is much shorter, so the variabiilty there is more easily
attributed to stellar rotation (St-Louis \etal\ 1995).  Hence there
remains the possibility that the concordance between the X-ray
variability timescale and the UV variability timescale might simply
reflect the coincidental matching of the flow time with the rotation
period.  However, if a more causally connected explanation is sought,
binarity could not operate on the necessary timescale.

For definitively ruling out the binary explanation, there
are some observational tests that would
prove useful.  First, WR~6 could be monitored in X-rays loosely at
monthly intervals over a period of years to look for cyclic trends,
rather than stochastic variations.
Second, as it happens, the radio photosphere of WR~6's wind is
roughly similar in radial scale to the radius of optical depth unity
for X-ray photoabsorption.  The presence of a binary companion
separated from WR~6 at a similar length scale should modify the
effective shape of that photosphere, and so might cause modulation
of the radio continuum light curve that would be correlated with
X-ray variations.  Both of these tests would require sparse, but
dedicated, long-term monitoring of WR~6.

Here we explore the possibility of a non-binary explanation that unifies
the 3.766~d period seen in other wavebands with the production of X-rays
at large radius.  Co-rotating interaction regions (CIRs) were proposed to
explain a variety of observed phenomena in the solar wind and have long
been invoked as a possible explanation for discrete absorption features
observed in the UV lines of OB stars (e.g., Mullan 1984; Hamann \etal\
2001), including WR~6 (St-Louis \etal\ 1995).  A CIR arises from the
interaction of wind flows that have different speeds: rotation of the
star ultimately leads to a collision interaction between the different
speed flows to produce a spiral pattern in the wind.

Recent new work has suggested the presence of a CIR in WR~6
and a few other WR stars showing similar variability in the optical
(St-Louis \etal\ 2009; Chene \etal\ 2011; Chene \& St-Louis 2011).
We next consider a heuristic kinematic model of a CIR structure in the wind
of WR~6 associated with the 3.766~d period in the form of
stellar rotation.

\section{Applying a CIR Model to the X-ray Variability}
\label{sec:model}

The motivation for the CIR is the idea that wind shocks degrade over
time (Gayley 2013), and since the X-ray emission from WR~6 arises from
large radii of $\sim 10^2-10^3 R_\ast$, it is difficult to understand how
the wind-shock paradigm could account for the presence of hot plasma.
A CIR represents a globally ordered pattern that might conceivably
persist to large radius, as seen in systems like the ``dusty pinwheels''
for WR binaries (e.g., Tuthill, Monnier, \& Danchi 1999; Monnier,
Tuthill, \& Danchi 1999; Harries \etal\ 2004; Tuthill \etal\ 2008).
In the application here, the CIR is associated with a single star and
related to stellar rotation, not orbital revolution.

In what follows a description of a model for an equatorial 
CIR region is presented.  A key assumption is that
the observed X-ray emission forms at large radius in the flow owing
to strong photoabsorption by the the dense WR~wind.  At large radii
the wind mass density is approximately

\begin{equation}
\rho = \frac{\dot{M}}{4\pi\,r^2\,v_\infty} = \rho_0\,u^2,
	\label{eq:rho}
\end{equation}

\noindent where the inverse radius $u=R_\ast/r$ is convenient
to use in this analysis.  The number density of
electrons is then given by $n_{\rm e} = n_0\,u^2$, with
$n_0 \propto \rho_0\propto \dot{M}/v_\infty$.

Before introducing the CIR structure, it is first useful to review
results for an otherwise smooth spherial wind as providing a
background against which to interpret the influence of a CIR for
variable X-ray emissions.  

\subsection{The Solution for a Smooth, Spherical Wind}

Models for the X-ray generation produced throughout a wind have
been discussed by numerous authors, and often parameterized in terms
of a smooth, spherical wind flow with a volume filling factor of
hot X-ray emitting plasma (e.g., Baum \etal\ 1992;
Hillier \etal\ 1993; Owocki
\& Cohen 1999; Oskinova \etal\ 2001).  Although the clumped nature of
massive star winds has long been recognized, considerations of the
smooth wind case has value in its simplicity and continues to provide
broad insight into the main features of how a distribution of X-ray
sources in a wind combine with wind photoabsorption effects to
create an emergent X-ray spectrum.  We consider a smooth wind
model before turning to the more complex CIR case.

Imagine a spherically symmetric and laminar wind flow.  Since the
X-rays emerge only from large radius, we restrict ourselves to the
use of equation~(\ref{eq:rho}) for an inverse square density law.
One of the key factors governing the emergent X-ray spectrum is the
wind photoabsorption.  The optical depth $\tau$ of photoabsorption
along a sightline through the spherical wind is given by

\begin{equation}
\tau = \int_z^\infty \, \kappa(E)\,\rho\,dz
\end{equation}

\noindent where $z$ is the coordinate along the sightline of the
observer (located at $+\infty$), and $\kappa(E)$ is the 
energy-dependent opacity for photoabsorption.  The optical depth of
photoabsorption has an analytic solution when $\rho \propto u^2$,
as given by (e.g., Ignace 2001):

\begin{eqnarray}
\tau(r,\theta, E) & = & \kappa(E)\,\rho_0\,R_\ast\times u\,\frac{\theta}
	{\sin\theta}, \\
 & \equiv & \tau_0(E)\,u\,\frac{\theta}{\sin\theta},
\end{eqnarray}

\noindent where $\theta$ is the polar angle from the observer's
axis, and $\tau_0(E)$ is a parameterization for the energy dependence
of the optical depth.  

The luminosity of presumably optically thin X-ray emission formally
derives from a volume integral:

\begin{equation}
L(E) = \int dV \, \int dT \, {df \over dT} \,n^2(r)\,G(E,T)\,e^{-\tau(r,
\theta,E)},
\end{equation}

\noindent where $df/dT$ is the distribution of volume filling factor over $T$,
and $G(E,T)$ is the appropriate 
kernel for collisional ionization equilibrium
(neglecting density dependence) for distributing the X-ray
emission over photon energy $E$.
Making for simplicity the approximation that $df/dT$ does not depend
on $r$, we can separate the $r$-independent terms into a single 
``emissivity profile'' given by

\begin{equation}
\ef(E)  =   {1 \over f_V} \, \int dT \, {df \over dT} \,G(E,T),
\end{equation}

\noindent where $f_V = \int dT \, df/dT$
is the volume filling factor of hot gas, assumed constant.
This separation yields

\begin{equation}
L(E) = \ef(E) \int dV \,n^2(r)\,e^{-\tau(r,\theta,E)},
\end{equation}

\noindent where
the form of $\ef(E)$ can then be chosen separately to mimic the actual spectrum.
The interest here is on how the
escape physics affects the total luminosity; the as-yet unspecified
processes that shape the intrinsic emissivity profile $\ef(E)$ can
be addressed in future studies.

Substituting the previous relationships into the integrand, and
evaluating both the azimuthal and radial integrations, the spectral energy
distribution in the luminous output becomes

\begin{eqnarray}
L(E) & = & 2\pi\,R_\ast^3\, f_V\,\frac{\ef(E)}{\tau_0(E)} \times \nonumber \\
 &  & \int_0^\pi\left[ 1 - e^{-\tau_0\,\theta/\sin\theta}\right] 
	\frac{\sin^2\theta}{\theta} d\theta.
\end{eqnarray}

\noindent In the limit that $\tau_0(E) \gg 1$, the second term in the
brackets of the integrand vanishes, and the integral yields a constant
value of 1.2188.  In this case the luminous spectrum is

\begin{equation}
L(E) \propto \frac{\ef(E)}{\tau_0(E)}.
\end{equation}

\noindent Note that in this approach, and with $f_V$ a constant,
the emergent spectral energy distribution (SED) is given strictly
by the ratio of the energy-dependent profile function to the
energy-dependent photoabsorptive opacity.

\subsection{A CIR X-ray Source Model}

To produce a reabsorbing environment that can show rotational
modulation, we next introduce a simple CIR model in the form of a
spiral pattern in the flow, following Ignace, Hubrig, \& Sch\"{o}ller
(2009) and Ignace, Bessey, \& Price (2009).  For a fixed radius,
the CIR is taken to have a circular cross-section.  The opening
angle of this cross-section is denoted as $\gamma$.  The equation
of motion for the center of the spiral, in the limit that $r \gg
R_\ast$, is given by

\begin{equation}
\varphi_{\rm c}(r) = \varphi_0 + \omega\, t - \frac{r\,\omega}{v_\infty}, 
\end{equation}

\noindent where $\omega = 2\pi / P_{\rm rot}$, with $P_{\rm rot}$
the stellar rotation period, and $\varphi_{\rm c}$ the azimuth
of the center as a function of radius.  A characteristic length
scale of this prescription is the ``winding radius'' given by

\begin{equation}
r_{\rm w} = v_\infty\,P_{\rm rot},
\end{equation}

\noindent which represents the length traveled by the wind in one
rotation period.  Consequently, it is related to the asymptotic
pitch angle of the spiral.

We assume that the CIR is the only source of X-ray emission at large
radius in the wind.  We are unaware of any calculations that would
provide guidance as to the density and temperature distribution in
such a structure at such distances.  For simplicity, and in order
to determine the potential plausibility of such a model, we will
assume that the density of the hot plasma scales with the wind
density (i.e., $r^{-2}$).  We further adopt a hot-plasma emissivity
profile $\ef$ with the form of a power law in energy and a low-energy
cut-off.  The dominant form of cooling for hot plasma at the
temperatures of interest is by lines.  However, there are a great
many weak lines in the spectrum, punctuated by several strong ones.
For a low-resolution SED, sufficient for our exploratory model, the
emission lines blend to form a pseudo-continuum.  The adopted
power-law form is roughly consistent with a multi-temperature plasma
(e.g., Owocki \& Cohen 2001); implicit is that the range of
temperatures and the relative amount of emission measure per
temperature interval are constants throughout the CIR.

So, the monochromatic luminosity is taken to be of the form:

\begin{equation}
L(E) = \ef(E) \, \int\,n^2(r)\,e^{-\tau(r,\theta,E)}\,dV.
\end{equation}

\noindent For the profile function, we adopt the following:

\begin{equation}
\ef(E) = \ef_0\, E^{-q_2}\,\left\{1-\exp\left[\left(E/E_0\right)^{q_1}\right]\right\},
\end{equation}

\noindent where the observed X-ray spectrum (see Fig.~\ref{fig1})
guides the selection of the constants $q_1$, $q_2$, and $E_0$.  In
our treatment only the CIR emits X-rays at the large radii of
interests.  X-rays that emit in the direction of the observer are
still attenuated by the wind.  To evaluate this absorption, we adopt
a smooth wind for these purposes, for which the optical depth will
be given by

\begin{equation}
\tau = \tau_0(E)\,\frac{R_\ast}{r}\,\frac{\theta_{\rm c}}{\sin\theta_{\rm c}},
\end{equation}

\noindent where

\begin{equation}
\tau_0(E) = \kappa(E)\,\rho_0\,R_\ast,
\end{equation}

\noindent with 

\begin{equation}
\rho_0 = \frac{\dot{M}}{4\pi\,R_\ast^2\,v_\infty},
\end{equation}

\noindent and

\begin{equation}
\kappa(E) = \kappa_0\,\left(E/1~{\rm keV}\right)^{-Q},
\end{equation}

\noindent with $Q$ a constant.  In our models the constants $\kappa_0$
and $\ef_0$ are set to unity as we mainly seek to reproduce the
overall shape of the observed X-ray spectrum.  The remaining
parameters are chosen to accomplish this end, with values of $E_0
= 0.6$~keV, $q_1 = 2.8$, $q_2 = 3.8$, and $Q = 2.6$.

We assume that the CIR has a relatively small opening angle, so
that the optical depth to the center of the structure at radius $r$
represents the column of absorbing material to the CIR's entire
cross-section at that radius. Then the resultant energy-dependent
luminosity reduces to

\begin{equation}
L(E) = L_0(E)\,\int_0^1\,\tau_0(E)\,e^{-\tau_0(E)\,u\,
	\theta_{\rm c}/\sin\theta_{\rm c}}\,du,
	\label{eq:genCIR}
\end{equation}

\noindent where again $u=R_\ast/r$, and

\begin{equation}
L_0(E) = \pi\,\gamma^2\,n_0^2\,R_\ast^3\,\frac{\ef(E)}{\tau_0(E)}.
\end{equation}

\noindent The resultant X-ray spectrum separates into a pair of
factors, one of which mimics the result of the smooth, spherical wind
in the form of $L_0(E)$.  The other factor is the implicit
energy-dependent, and time-dependent, integration owing to the
CIR structure.

The integral expression in equation~(\ref{eq:genCIR}) does not have
a general analytic solution given that $\theta_{\rm c} = \theta_{\rm c}(u)$.
However, there is a special case where a solution can be derived
and which provides some insight for application to the observations
of WR~6, even though it is not strictly appropriate.  
Consider the case of a long, straight CIR, which is the limit
of $v_{\rm rot} \ll v_\infty$.  Taking into account the photoabsorption
of X-rays by the wind, ``long'' implies that $r_1 \lesssim r_{\rm w}$
such that the CIR is effectively a cone over the relevant region of emission.
In this limit the solution for the SED becomes

\begin{equation}
L(E) \propto \frac{\ef(E)}{\tau_0(E)} \times \frac{\sin\theta_{\rm c}(t)}
	{\theta_{\rm c}(t)},
\end{equation}

\noindent where from spherical trigonometric considerations

\begin{equation}
\cos \theta_{\rm c} = \sin i\,\cos \varphi_{\rm c}(t) = \sin i\,\cos(\Omega\,t).
\end{equation}

\noindent In this limiting case, the viewing inclination sets the
maximum amplitude of variation.  But what is clearly evident is the
cyclic nature of the variation.  Allowing for the spiral nature of
the CIR does not change the fact that the model produces cyclic
variations at fixed $E$.  However, relaxing the straight-arm cone
to a spiral pattern does introduce phase shifts in the cyclic
variability from one energy to the next.  

A phase lag can be understood qualitatively by thinking in terms
of how the spiral CIR intersects a sphere of the energy-dependent
radius of unit optical in photoabsorption, $r_1(E)$.  That intersection
occurs at different azimuths for different energies because of the
fact that $r_1$ is larger at softer energies and smaller at harder
ones.  Indeed, the azimuthal location of the CIR's center at radius
$r_1$ is given by $\varphi_1 = \varphi_{\rm c}(t,r_1) = \varphi_{\rm
c}(t,E)$.  Even so, X-ray bandpass count rates or fluxes will {\it
still} be cyclic on the period of the stellar rotation.  Additionally,
because of the energy-dependent phase lags, the overall effect will
be to {\em reduce} the amplitude of variation in a bandpass as
compared to the monochromatic variability.  Obtaining cycle-to-cycle
variations requires a new ingredient to the model.

\begin{figure}[t!]
\centering
\includegraphics[width=1.0\columnwidth]{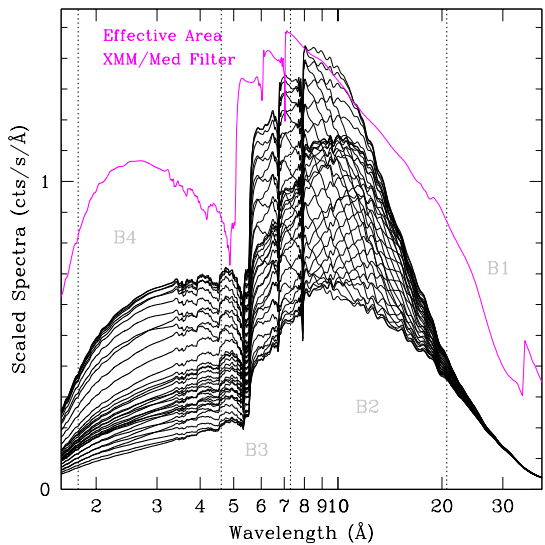}
\caption{Linear-log overplot of {\em synthetic}
XMM-EPIC spectra (PN+MOS1+MOS2) for the
medium filter (area response in magenta) illustrating the variable
SED arising from the CIR model.  Vertical dotted lines in gray
signify the bands B1--B4 used in this paper.
} 
\label{fig7} 
\end{figure}

\begin{figure}[t!]
\centering
\includegraphics[width=1.0\columnwidth]{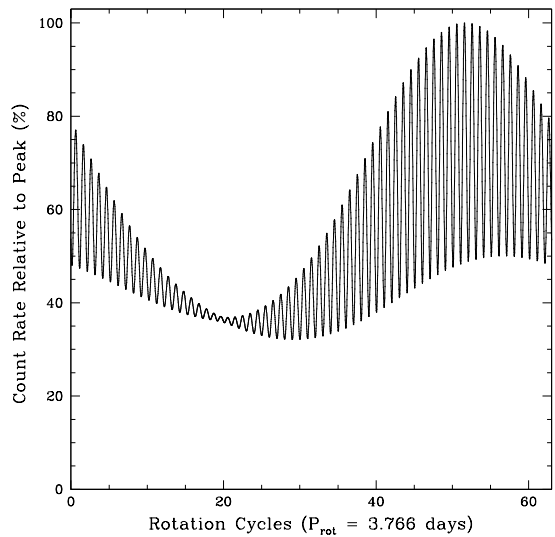}
\caption{An example light curve for the total X-ray count rate
$\dot{\cal C}_{\rm T}$ from our ``wavy'' CIR model.  The count rate
is normalized to have a peak value of unity; here the count rate
is displayed as a percentage of the peak value.  The abscissa is
for cycles in terms of the 3.766~d period of WR~6.  The model
parameters in this exampe are for Model 17 from Tab.~\ref{tab3}.
}
\label{fig8} 
\end{figure}

\begin{figure}[t!]
\centering
\includegraphics[width=1.0\columnwidth]{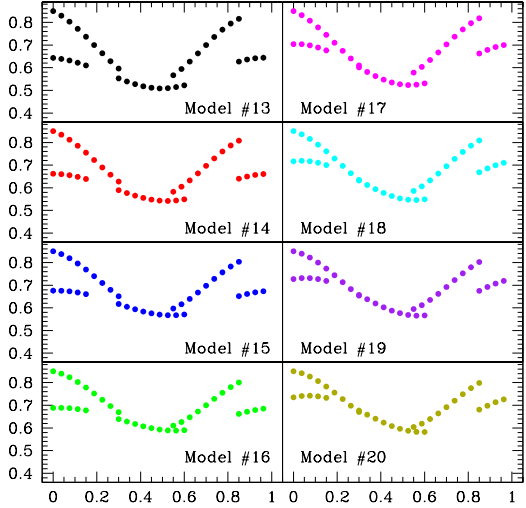}
\caption{Example phased light curve with the modulated CIR model.
These segments correspond to the observed phases of our
four pointings.  The parameters for these light curves are
for models 13 to 20 from Table~\ref{tab3}.  All of
the model light curves were scaled to have peak values similar
to the observed count rates.} 
\label{fig9} 
\end{figure}

\subsubsection{The Wavy CIR}

The discussion of the preceding section for a steady-state CIR yielded
a variable X-ray signal because of the wind photoabsorption in relation
to the rotating and non-axisymmetric structure of the CIR.  A CIR is
a hydrodynamic phenomenon.  It may be possible that its structure is
influenced by hydrodynamic instabilities.  CIR structures from the
solar wind are fairly well-studied (e.g., Rouillard \etal\ 2008), and it
is known that CIRs can merge at large distances in the solar wind, at
around 10--20~AU (Burlaga, Schwenn, \& Rosenbauer 1983; Burlaga, Ness,
\& Belcher 1997).  Moreover, the large-scale solar magnetic field is
dipolar but tilted somewhat to the rotation axis.  This magnetic field
can interact with a CIR leading to a ``deflection'' of the structure,
meaning for example that spiral path of the CIR no longer resides in a
single plane (Gosling \& Pizzo 1999).

Merging and deflection of solar CIRs are complex effects, and it
is difficult to speculate how such behavior in the solar wind might
carry over to the highly unstable massive-star winds.  Nonetheless, the
case of the Sun informs us that there are processes that can modify CIR
structures on time scales that are unrelated to the rotation of the star.
Instabilities associated with line-driven wind theory or effects like
those seen with solar CIRs can plausibly lead to variable structure
in the CIR itself with consequent cycle-to-cycle (or epoch-to-epoch)
variations in the X-ray emissions.

In an attempt to understand the X-ray variations of WR~6, we introduce
a simple modification to the CIR structure.  Given that one would
not expect the rotation of the star to vary, there are two main
options for our model: either the opening angle of the CIR or the
wind terminal speed changes in time.  The wind terminal speed would
govern the pitch angle of the spiral as a function of time and
location; the opening angle $\gamma$ would govern the solid angle
of the CIR as a function of time and location.  For purposes of
illustration, we choose to allow for a variable opening angle of
the CIR, which we refer to as the ``wavy'' CIR.

To accomplish this, we imagine a sinusoid wave propagating along
the length of the CIR.  We model the opening angle with

\begin{equation}
\gamma = \gamma_0 \, \left[ 1 + \delta \,\sin(Kr-\Omega t)\right],
\end{equation}

\noindent indicating that the opening angle varies from $(1-\delta)
\gamma_0$ to $(1+\delta)\gamma_0$.  The wave parameters $K$ and
$\Omega$ are related to the characteristic length and the period
of the wave as $l_{\rm wav} = 2\pi/K$ and $P_{\rm wav} = 2\pi/\Omega$.
The wave travels along the CIR at speed $v_{\rm CIR} = \Omega/K$.  The
combination of the wind photoabsorption with a non-axisymmetric
spiral pattern and an additional time-varying CIR structure that
is uncorrelated with the stellar rotation yields a generally cyclical
X-ray light curve with an overlying cycle-to-cycle modulation.

\begin{figure}[t!]
\centering
\includegraphics[width=1.0\columnwidth]{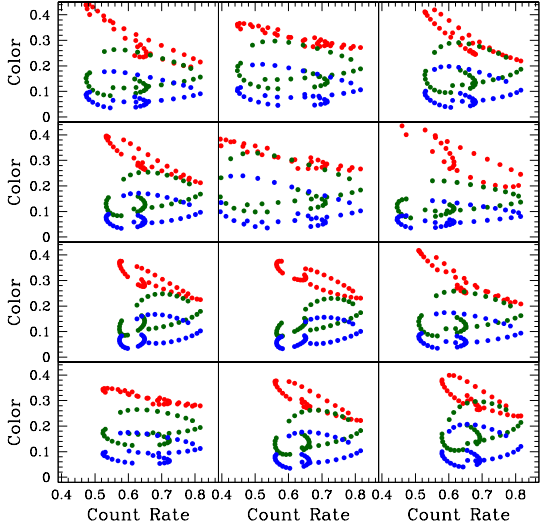}
\caption{Models 1 to 12 as parameter study plotted as X-ray color
versus count rate to mimic Fig.~\ref{fig6}.  The color scheme is the
same as for Fig.~\ref{fig6}, with red for Soft, green for Medium, and
blue for Hard.  The same relative phases are shown here as for the
observations, but with a lower density of points.  Model parameters
are given in Tab.~\ref{tab3}.  From top to bottom, the panels are for
models 1--4 at left, 5--8 in the middle, and 9--12 at right.
} 
\label{fig10} 
\end{figure}

\begin{figure}[t!]
\centering
\includegraphics[width=1.0\columnwidth]{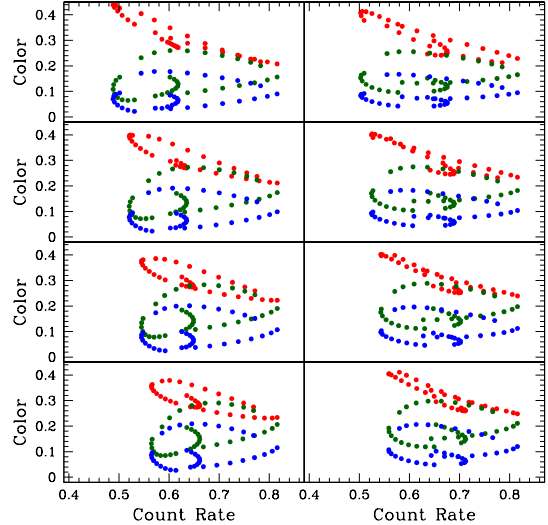}
\caption{Models 13 to 20 with variations geared to match more
closely those observed in WR~6.  See Tab.~\ref{tab3} for the parameters.
In particular the ratio $v_{\rm rot}/v_\infty$ is fixed at 0.009 as the
best estimate for WR~6.  Again, the color scheme mimics Fig.~\ref{fig6}.
Left panels are models 13--16 from top to bottom; at right the models
are 17--20.}
\label{fig11} 
\end{figure}

In an exploration of model results, we have calculated a number of
example models as shown in Figures \ref{fig7}--\ref{fig11}.  Parameters
used for these models are listed in Table~\ref{tab3}.  Note that in
this table, the wave parameters $\Omega$ and $K$ are given in terms
of the stellar rotation period (in this case $P_{\rm rot} = 3.766$~d)
and the stellar radius.  For example, all of the models use
$\Omega = 0.1/P_{\rm rot}$, meaning that the period of the propagating
wave for the CIR is $P_{\rm wav} = 2\pi/\Omega = 20\pi\,P_{\rm rot}$.

We have not attempted to reproduce exactly the observed variations
shown in Figure~\ref{fig5}.  Although the spiral pattern is physically
motivated, the forms for the wavy CIR and the X-ray emission
distributions are merely convenient prescriptions.  Consequently,
it is premature to attempt any quantitative fits to the data, and
instead our focus is on the qualitative aspects of a CIR that might
account for the observed variability.

Figure~\ref{fig7} shows the variable spectrum, in counts/s/\AA,  as
a function of time.  The black curves are model spectra at different
times as the star rotates.  Indicated in gray are the four energy
bands used in our analysis.  Also shown in magenta is the total
{\em XMM-Newton} effective area, using a medium filter, as taken
from the PIMMS software (Mukai 1993).  The variations are relative
to a reference spectrum of unit area.

An example of the long-term variability that can arise from a ``wavy
CIR'' is displayed in Figure~\ref{fig8}.  The model parameters
correspond to 17 of Table~\ref{tab3}.  The time axis is in
terms of the number of cycles of the 3.776 day period for WR~6.
With $\Omega = 0.1$, the period of the CIR wave is $P_{\rm wav}/P_{\rm
rot} \approx 63$ rotational cycles.  The variable X-ray count rate
is displayed as a percentage of the peak value obtained.  The rapid
variations are of course the 3.776~d period.  The long term modulation
reflects the time- and location-dependent opening angle of the CIR.

Figure~\ref{fig9} illustrates how the wavy CIR leads to disjoint
phased light curves.  Each panel is a model light curve, with model
number indicated, phased on the 3.766~d period.  The vertical is
counts per second scaled to similar values as those observed.  The
segments correspond to the relative time durations and intervals
for our four pointings.  If $K$ and $\Omega$ were zero, the light
curves would all be continuous; addition of a ``wave'' to the CIR
structure leads to epoch-dependent variations in the total count
rates.

Figure~\ref{fig10} shows a grid of model results for relative colors
versus total count rates analogous to Figure~\ref{fig6}.  The red,
green, and blue points are the same as in Figure~\ref{fig6}.  The
total count rate is simply a scaling applied to the model results
to match roughly the maximum count rate obtained in the observations;
colors have the advantage that the model results are independent of
the scaling used.

Model parameters for this grid are given in Table~\ref{tab3}
for models labeled 1--12.  The units of $K$ and $\Omega$ are such
that the wave speed is given by $v_{\rm CIR} = 2.15~{\rm km/s} \times
K/\Omega$ with $R_\ast = R_\odot$ for WR~6 and $P_{\rm rot} =
3.766$~d.  Similarly, values for the ratio $v_{\rm rot}/v_\infty$
relevant to WR~6 are motivated by the observed wind terminal speed
of 1600 km/s.  The optical depth scale $\tau_0$ at 1~keV is simiarly
motivated by the case of WR~6.  The key point is that for the phases
of the pointings for WR~6, Figure~\ref{fig10} shows that variations
in the relative color and total count rate do result.

Figure~\ref{fig11} is similar to Figure~\ref{fig10}, except that
$v_{\rm rot}/v_\infty=0.009$ is fixed to match our best estimate
for this ratio.  Parameters for these calculations correspond to
models 13--20 in Table~\ref{tab3}.  With $v_{\rm rot}/v_\infty$,
we show models for a greater range of $\tau_0$ values.  The first
four models are for higher inclination perspectives as shown at
left; the last four models are perspectives for a mid-latitude view
shown in the right column of panels.  Note that variability does
result for a pole-on view because of the time-dependent opening
angle of the CIR; no variability would result for a pole-on view
if $\Omega=0$.

\begin{table*}[t]
\begin{center}
\caption{Model Parameters \label{tab3}}
\begin{tabular}{cccccccc}
\hline\hline Model & $\Omega$ & $K$ & $\delta$ & $\gamma_0$ & $i$ & $v_{\rm rot}/v_\infty$ & $\tau_0$ \\ 
   & $(P_{\rm rot}^{-1})$ & $(R_\ast^{-1})$ &  & (degs) & (degs) & & (@ 1 keV) \\ \hline
 1 & 0.10 & 0.005 & 0.5 & 10 & 50 & 0.006 & 200	\\
 2 & 0.10 & 0.005 & 0.5 & 10 & 50 & 0.008 & 200	\\
 3 & 0.10 & 0.005 & 0.5 & 10 & 50 & 0.010 & 200	\\
 4 & 0.10 & 0.010 & 0.5 & 10 & 30 & 0.010 & 200	\\
 5 & 0.10 & 0.010 & 0.5 & 10 & 45 & 0.010 & 200	\\
 6 & 0.10 & 0.010 & 0.5 & 10 & 60 & 0.010 & 200	\\
 7 & 0.10 & 0.004 & 0.5 & 10 & 50 & 0.010 & 200	\\
 8 & 0.10 & 0.006 & 0.5 & 10 & 50 & 0.010 & 200	\\
 9 & 0.10 & 0.008 & 0.5 & 10 & 50 & 0.010 & 200	\\
10 & 0.10 & 0.005 & 0.5 & 10 & 50 & 0.007 & 100	\\
11 & 0.10 & 0.005 & 0.5 & 10 & 50 & 0.007 & 200	\\
12 & 0.10 & 0.005 & 0.5 & 10 & 50 & 0.007 & 400	\\ \hline
13 & 0.10 & 0.005 & 0.5 & 10 & 70 & 0.009 & 150	\\
14 & 0.10 & 0.005 & 0.5 & 10 & 70 & 0.009 & 200	\\
15 & 0.10 & 0.005 & 0.5 & 10 & 70 & 0.009 & 250	\\
16 & 0.10 & 0.005 & 0.5 & 10 & 70 & 0.009 & 300	\\
17 & 0.10 & 0.008 & 0.5 & 10 & 40 & 0.009 & 150	\\
18 & 0.10 & 0.008 & 0.5 & 10 & 40 & 0.009 & 200	\\
19 & 0.10 & 0.008 & 0.5 & 10 & 40 & 0.009 & 250	\\
20 & 0.10 & 0.008 & 0.5 & 10 & 40 & 0.009 & 300	\\ \hline
\end{tabular}
\end{center}
\end{table*}

\section{Conclusions}
\label{sec:conc}

A remarkable dataset of WR~6 was obtained by the {\em XMM-Newton}
telescope, amounting to four pointings of approximately 100~ks each.
The pointings came in the form of ``on'' and ``off'' exposures of
two pairs of day-on, day-off, and day-on sequences, with the pairs
separated by a few weeks.  The high quality data provide a unique
opportunity to probe the wind of WR~6 to understand better how its
high energy emissions are produced and to understand the well-known
3.766~d variability period that has been observed in many other
wavebands.

The four pointings cover nearly all phases of the 3.766~d period, with some
overlap as well.
Although variability is clearly evident, we were somewhat surprised
to find that star's signature 3.766~d period is not obviously present
in the observations.  The dataset is of sufficient quality that we
could conduct a search for ``event clustering'' in the arrival times
of individual X-ray photons, but we fail to detect a signature of
temporal clustering in the photon counts.  The absence of such
clustering would tend to favor either a tremendously large number
of clumps or a stationary shock structure.  The former would be
consistent with the interpretation of variability seen in the O~star
$\zeta$~Pup.  The latter would be more consistent with a global
wind structure.  However, it seems physically quite challenging to
understand how a stochastically structured wind, like that predicted
by the LDI mechanism, could produce hot gas at large radii in the
wind as implied by the spectral analysis of Paper~I.  In other words
it is difficult to imagine how velocity differences of 100's of
km/s, required to produce X-rays at 1~keV energies, could exist at
100's of stellar radii, well beyond the zone where the wind is
accelerated to terminal velocity.

Discarding the embedded wind shock paradigm leaves the case of a
stationary shock. The two most obvious candidates for such a structure
would be a binary colliding wind interaction or a CIR feature.  We have
argued against the binary hypothesis in favor of a CIR structure.

Unfortunately, the large-scale coherence and evolution of a CIR in
the context of a WR~wind has not previously been investigated, and
the nature of the X-ray production is unknown.  Consequently, we
considered a simplistic kinematic model of a CIR as a spiral feature
that threads the wind, and we used scaling arguments to motivate a
qualitative approach to the X-ray spectral energy distribution as
a function of radius in the wind.  Our key motivation has been
simply to explore the qualitative behavior of such a model.  We
find that a CIR does provide features that could explain the observed
variability in WR~6.  However, in order to accommodate the lack of
precisely cyclical behavior in the X-ray light curve, a perturbation
of the CIR structure was invoked.  Specifically, we allowed for
propagation of a wave along the length of the CIR that served to
modulate the opening angle (and thus emission measure) of the spiral
structure.

It is not difficult to imagine instabilities that might serve to
drive such a result.  Certainly, the increasing evidence in support
of CIRs among WR~stars (e.g., St-Louis \etal\ 2009) indicates a
need to explore the X-ray signatures that could result from such
structures for WR~winds.  We also recognize deficiencies in
our model, such as implicitly treating the photoabsorbing
opacity as a constant with radius whereas there is evidence to
the contrary for massive star winds (e.g., Oskinova \etal\ 2006;
Herv\'{e} \etal\ 2012). 

We suggest that the next step in understanding the X-ray processes in
operation in WR~6 will require a long-term ``spot check'' monitoring
program.  X-ray count rates obtained at roughly weekly intervals
over the course of about a year at a detection S/N of roughly 20
should be adequate to determine whether the X-rays vary at the
3.766~d period with an additional modulation like that indicated with
our ``wavy CIR'' model.

\section*{Acknowledgements}

The authors express appreciation to an anonymous referee for
several useful suggestions.
DPH was supported by NASA through the Smithsonian Astrophysical
Observatory contract SV3-73016 for the Chandra X-Ray Center and
Science Instruments. Funding for this research has been provided
by DLR grant 50~OR~1302 (LMO).

\end{document}